\documentclass[global,twocolumn]{svjour}

\usepackage{graphicx}
\usepackage{mathptmx} 
\usepackage{multirow}
\usepackage{siunitx}
\usepackage{subfigure}

\journalname{Applied Physics B}

\begin{document}

\title{Single photon emitters based on Ni/Si related defects in single crystalline diamond}

\author{David Steinmetz\inst{1}\and Elke Neu\inst{1}\and Jan Meijer\inst{2}\and Wolfgang Bolse\inst{3}\and Christoph
   Becher\inst{1}\thanks{Fax: +49-681-302-4676,\newline E-Mail: christoph.becher@physik.uni-saarland.de}}

\institute{Universit\"at des Saarlandes, Fachrichtung 7.3 (Technische Physik), Campus E2.6, 66123 Saarbr\"ucken, Germany
   \and RUBION, Ruhr-Universit\"at Bochum, Universit\"atsstra{\ss}e 150, 44780 Bochum, Germany
   \and Universit\"at Stuttgart, Institut f\"ur Halbleiteroptik und Funktionelle Grenzfl\"achen, Allmandring 3, 70569 Stuttgart,
      Germany}

\date{Received: / Revised version:}

\maketitle

\begin{abstract}
We present investigations on single Ni/Si related color centers produced via ion implantation into single crystalline type IIa CVD diamond. By testing different ion dose\linebreak combinations we show that there is an upper limit for both the Ni and the Si dose (\SI{E12}{cm^{-2}} and \SI{E10}{cm^{-2}} resp.) due to creation of excess fluorescent background. We demonstrate creation of Ni/Si related centers showing emission in the spectral range between \SI{767}{nm} and \SI{775}{nm} and narrow line-widths of $\approx\SI{2}{nm}$ FWHM at room temperature. Measurements of the intensity auto-correlation functions prove single-photon emission. The investigated color centers can be coarsely divided into two groups: Drawing from photon statistics and the degree of polarization in excitation and emission we find that some color centers behave as two-level, single-dipole systems whereas other centers exhibit three levels and contributions from two orthogonal dipoles. In addition, some color centers feature stable and bright emission with saturation count rates up to \SI{78}{kcounts/s} whereas others show fluctuating count\linebreak rates and three-level blinking.\\[0.5cm]
\textbf{PACS} 42.50.Ar, 81.05.ug, 61.72.jn
\end{abstract}

\section{Introduction}

Single-photon sources are of great interest because of the numerous possible applications in quantum information processing.
A prominent example is the BB84-protocol for quantum key distribution \cite{Bennett1984}  that can be realized with single photons as information carrier. For this application one requires a single-photon source that can be triggered, features a high emission rate to assure fast information transmission and exhibits a narrow line-width to allow for an effective spectral signal filtering. In principle, such a source can be realized by a single emitting quantum system. If such a dipole undergoes cycles of pulsed excitation, light emission and re-excitation before emitting the next photon, triggered single-photon emission can be achieved. Admittedly, many of the existing demonstrations e.g. single atoms \cite{Kuhn2002} or single quantum dots \cite{Santori2001} hold experimental complexity that limits their suitability for practical applications.

In comparison, color centers in diamond have larger potential for feasible single-photon sources: These point defects consisting of combinations of impurity atoms and lattice vacancies exhibit an atom like spectrum even at room temperature (RT)
\cite{Zaitsev2001}. Additionally the high photo-stability of many color centers offers the great advantage of long-time stable single-photon emission \cite{Lounis2005}.

The best known color center is the so called Nitrogen-Vacan\-cy center (NV) with a zero-phonon line (ZPL) at \SI{638}{} nm. Beyond single-photon emission \cite{Kurtsiefer2000} several key experiments towards quantum information have already been performed, including e.g. the realization of a two-qubit gate \cite{Jelezko2004a} and multipartite entanglement \cite{Neumann2008}. However, NV centers have the detrimental property of a broad emission bandwidth of $\approx\SI{100}{nm}$ due to strong vibronic coupling to the diamond lattice. This fact strongly limits their suitability for applications in quantum optics and quantum information where narrow-band emission of single photons is essential, e.g. quantum key distribution or linear optics quantum computation.

A more auspicious kind of color centers are nickel-related centers. NE8 centers, defects that consist of one Ni atom and four N atoms \cite{Nadolinny1999}, show a narrow line (\SI{1.2}{nm} FWHM) around \SI{800}{nm} at RT \cite{Gaebel2004}. For single NE8 centers, photon anti-bun\-ching \cite{Gaebel2004} and triggered single-photon emission \cite{Wu2007} have been demonstrated. Investigations of color centers consisting of Ni and possibly Si in nano-crystals yielded high saturation count rates of \SI{200}{kcounts/s} at an emission wavelength of \SI{768}{nm} with an excited-state lifetime of \SI{2}{ns} \cite{Aharonovich2009}. However, targeted placing of single emitters for applications in quantum information requires their deterministic creation. Recent\-ly the first production of single Ni/Si related color centers via ion implantation in single crystalline diamonds has been reported \cite{Steinmetz2010}. We here present further investigations on optimized implantation parameters of Ni/Si centers and their optical and structural properties.

\section{Experimental}

For our investigations we used type IIa CVD single-crystal\-line diamonds (Element Six, Isle of Man, UK). Since the exact atomic composition and the creation efficiency of the Ni/Si related color centers are still unknown, we systematically varied implantation energies and doses of Ni and Si ions to find optimal implantation parameters. We are aiming at a color center density that is as high as possible and still allows for the localization of single centers with a confocal microscope. We chose two different implantation energies: at first to investigate the influence of the implantation energy on the creation efficiency and secondly to explore the dependence of emission properties on the distance of color centers from the diamond surface: On the one hand, if the distance of a single emitter from the diamond surface is larger than approximately half the emission wavelength, refraction at the surface becomes relevant, lowering the collection efficiency. On the other hand, surface effects might act detrimental on color centers if the distance to the surface is too small (e.g. NV centers change their preferred charge state depending on the distance to the surface \cite{Santori2009,Fu2010}). An overview of the implantation parameters can be found in the following table.
\begin{center}
   \vspace{0.5cm}
      \begin{tabular}{|c|c|c|c|}
      \hline
         ion & energy (keV) & depth (nm) & \phantom{\Large{I}}dose (ions/cm$^2$)\phantom{\Large{I}}\\
      \hline\hline
         Ni & 86 & \multirow{2}{*}{39} & \phantom{\Large{I}}\SI{E11}{}-\SI{E14}{}\phantom{\Large{I}}\\
         \cline{1-2}\cline{4-4}
         Si & 56 & & \phantom{\Large{I}}\SI{E9}{}-\SI{E12}{}\phantom{\Large{I}}\\
      \hline\hline
         Ni & 815 & \multirow{2}{*}{363} & \phantom{\Large{I}}\SI{E11}{}-\SI{E14}{}\phantom{\Large{I}}\\
         \cline{1-2}\cline{4-4}
         Si & 600 & & \phantom{\Large{I}}\SI{E9}{}-\SI{E12}{}\phantom{\Large{I}}\\
      \hline
      \end{tabular}
   \vspace{0.5cm}
\end{center}
The mean implantation depths were calculated with Monte Carlo simulations (SRIM\footnote{Stopping range of ions in matter: http://www.srim.org/}). Implantations were performed at the Georg-August Universit\"at (G\"ottingen, Germany) and at the RUBION (Ruhr-Universit\"at, Bochum, Germany). After the implantation, all samples were annealed at \SI{1200}{\celsius} in vacu\-um for one hour followed by cleaning in oxygen plasma to remove graphite from the sample surface formed during the annealing.

\begin{figure}\centering
   \includegraphics[width=7.5cm]{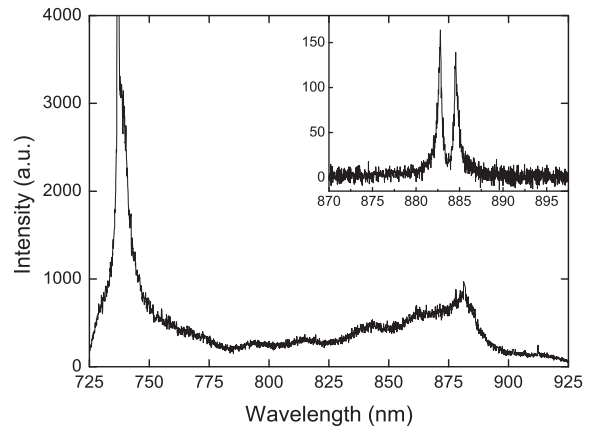}
      \caption{Emission spectrum of a sample implanted with Ni (\SI{86}{keV}, \SI{E13}{cm^{-2}}) and Si (\SI{56}{keV}, \SI{E11}
   {cm^{-2}}), excited with \SI  {671}{nm} laser light. Around \SI{738}{nm} the emission line of the SiV center is visible, the
   feature rising till \SI{881}{nm} is attributed to the so called \SI{1.4}{eV} center. Inset: detailed emission spectrum of the \SI{1.4}{eV} center ensemble measured at \SI{30}{K}.}
   \label{gesamtspektrum}
\end{figure}Fluorescence is excited through a microscope objective with $\text{NA}=0.8$ by a cw frequency doubled DPSS laser with a wavelength of \SI{671}{nm}. A set of dielectric filters blocks the excitation light from entering the analyzing beam path. The fluorescence can be guided into a grating spectro\-meter or into a Hanbury-Brown-Twiss setup (HBT) containing two ava\-lanche photo diodes (APDs) to measure the second order correlation function of the photoluminescence signal:
\begin{equation}
   g^{(2)}(\tau)=\frac{\langle I(t)I(t+\tau)\rangle}{\langle I(t)\rangle^2}\,.
\end{equation}
This function represents the probability to detect two photons separated by the time interval $\tau$. The intensity signal of the APDs can also be used for confocal scans of the diamond sample. Limited by the focus size of the excitation light a point like emitter appears with a lateral size of $\approx\SI{0.6}{\micro m}$ FWHM and an axial size of $\SI{1.9}{\micro m}$ FWHM in the scans. The complete detection efficien\-cy $\eta$ of our setup (taking into account the collection efficiency of the objective, the detection efficiency of the APDs and the transmission of all optical components) can be estimated as $\eta=\SI{2.2}{\%}$. All following measurements were performed at room temperature as far as not stated differently.

\section{Results and Discussion}

\subsection{Implantation Parameters}

\begin{figure}\centering
      \includegraphics[width=7.5cm]{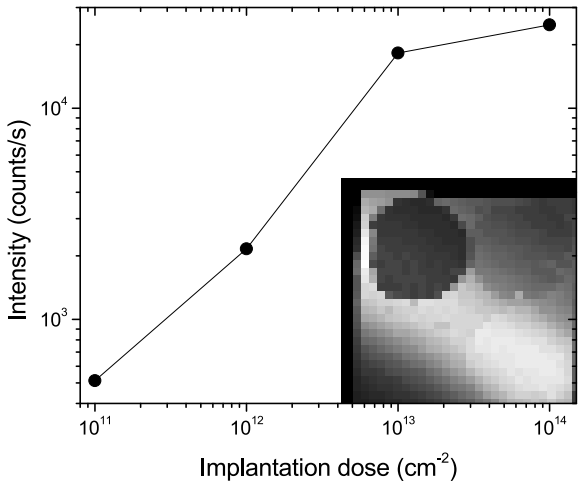}
      \caption{\SI{1.4}{eV} fluorescence intensity of implanted samples in dependence on the Ni implantation dose. Inset: SiV
      fluorescence on a sample homogeneously implanted with \SI{E12}{cm^{-2}} Si. Ni was implanted with four different
      doses in form of round circles in each corner. The dose decreases from \SI{E14}{cm^{-2}} to \SI{E11}{cm^{-2}}
      clockwise, starting in the upper left corner.}
   \label{ni_dosen}
\end{figure}
Photoluminescence investigations of samples implanted with high doses of both Ni and Si (\SI{E13}{cm^{-2}} and \SI{E11}{cm^{-2}}) essentially reveal two features in the room temperature fluorescence spectrum (see Fig. \ref{gesamtspektrum}) independent of the implantation energy: Around \SI{738}{nm} one can observe the ZPL of an ensemble of SiV centers (cf. \cite{Wang2006}). Following Sternschulte et al.  \cite{Sternschulte1994} the fluorescence up to \SI{825}{} nm can be attributed to the SiV vibronic sideband. The feature above \SI{825}{nm} with a maximum at $\approx\SI{881}{nm}$ is due to so called \SI{1.4}{eV} centers \cite{Collins1983}, possibly consisting of one Ni$^+$ ion at the center of a lattice di-vacancy \cite{Iakoubovskii2004}. At cryogenic temperatures, the emission spectrum of these color centers consists of two lines at \SI{883}{nm} and \SI{885}{nm} (see inset of Fig. \ref{gesamtspektrum} and Ref. \cite{Nazare1991}).

In Fig. \ref{ni_dosen} the intensity of the \SI{1.4}{eV} fluorescence measured on samples with different Ni doses is shown: The intensity follows a linear characteristic for the lower dose values as expected for a color center containing one Ni atom. It is obvious, however, that the slope reduces at the step between the two highest doses  \SI{E13}{cm^{-2}} and  \SI{E14}{cm^{-2}}. This behavior can be explained by increasing radiation damage influencing the creation efficiency of color centers and/or their emission properties negatively. The assumption of a damaged diamond lattice is supported by the observation of enhanced line widths of the two \SI{1.4}{eV} center lines  at a temperature of \SI{30}{K} with growing implantation dose: At a dose of \SI{E12}{cm^{-2}} we determined the widths of both lines of the doublet to be about \SI{0.3}{nm}, increasing up to \SI{0.7}{nm} at \SI{E14}{cm^{-2}}. This inhomogeneous broadening evidences a modified crystalline environment for single color centers.

The negative influence of too high Ni doses on the emission of other color centers can be revealed by a confocal overview scan of one of our samples (see inset in Fig. \ref{ni_dosen}). Although the sample is homogeneously implanted with Si ions, a reduced SiV ZPL emission intensity in the areas of the highest Ni densities is visible. These observations lead to an upper limit for the Ni implantation dose of \SI{E13}{cm^{-2}} to avoid similar emission quenching of single Ni/Si related centers.

\begin{figure}\centering
      \subfigure[Si: \SI{E9}{cm^{-2}}, Ni: see graph]{\includegraphics[width=5.5cm]{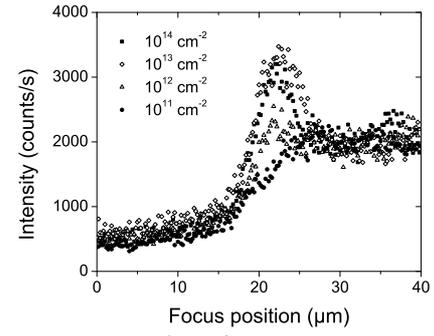}}\\[0.25cm]
      \subfigure[Ni: \SI{E11}{cm^{-2}}, Si: see graph]{\includegraphics[width=5.6cm]{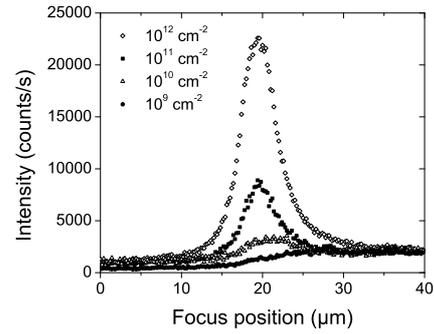}}
      \caption{Depth scans in the spectral window 765 -\SI{775}{nm} on eight different samples with different dose
      combinations.}
      \label{objektivscans}
\end{figure}

A further restriction for the ion dose is given by unwanted fluorescent background in the spectral range between \SI{760}{nm} and \SI{820}{nm} where we expect the Ni/Si related color center emission. This luminescence is produced by the vibronic features of the SiV centers and the fluorescence of the \SI{1.4}{eV} centers. Axial focus scans of different samples show a strong dependence of the unwanted fluorescence on the implantation doses (see Fig. \ref{objektivscans}) -- due to the strong SiV vibronic sideband, the effect is much stronger for SiV centers than for the \SI{1.4}{eV} centers. Thus, we determine maximum implantation doses \SI{E12}{cm^{-2}} for Ni and \SI{E10}{cm^{-2}} for Si to keep the unwanted fluorescence background at a tolerable level.

\subsection{Spectroscopy \& Single Photon Emission}

\begin{figure}\centering
      \includegraphics[width=7cm]{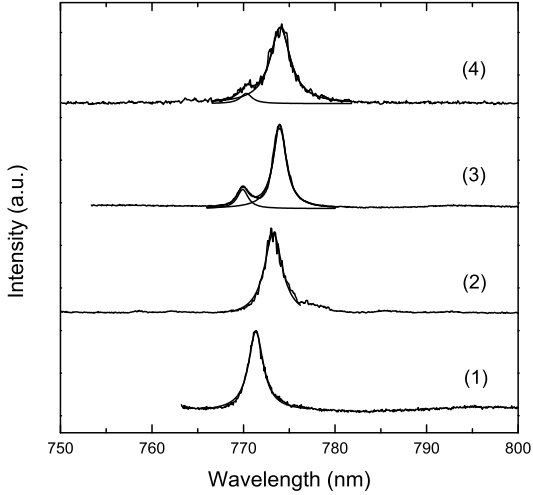}
      \caption{Emission spectra of four Ni/Si related color centers in the spectral range 765 -\SI{775}{nm}. The spectra can be
      fitted with one or two Lorentz peak functions. The line widths of the single peaks amount to 1.4\,-\,\SI{2.7}{nm}
      FWHM.}
      \label{spektren_vergleich}
\end{figure}

For the investigation of color center formation we performed confocal scans of our samples, detecting the fluorescence intensity in the spectral window of \SI{765}{nm}-\SI{820}{nm}. Although we were able to observe several color centers with emission wavelengths varying between 786 and \SI{815}{nm}, the majority of detected color centers showed an emission wavelength between 767 and \SI{775}{nm} (examples see Fig. \ref{spektren_vergleich}). The fact that these color centers can only be found in samples implanted with both Ni and Si (see Ref. \cite{Steinmetz2010}) leads, consistent with Aharonovich et al.
\cite{Aharonovich2009}, to the assumption that this last group are Ni/Si related color centers. However, the density of these color centers in our samples is still quite low: on average only one center per \SI{2000}{\micro m^2}, measured on samples with the optimal doses as discussed above (\SI{E12}{cm^{-2}} Ni and \SI{E10}{cm^{-2}} Si). This corresponds to creation efficiencies of \SI{5E-8}{} in relation to the Ni dose and \SI{5e-6}{} in relation to the Si dose. Due to these low values it was not possible to analyze the color center density in dependence on the ion implantation dose. The different implantation energies and resulting depths  of color centers did not lead to a significant difference in creation efficiency.

\begin{figure}\centering
      \subfigure[Emitter (2)]{\includegraphics[width=5.5cm]{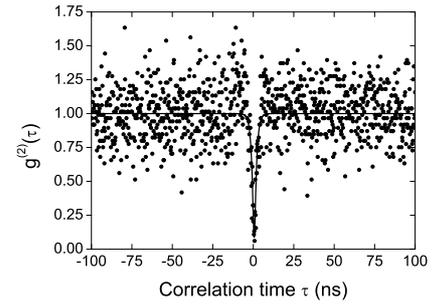}}\\[0.35cm]
      \subfigure[Emitter (4)]{\includegraphics[width=5.5cm]{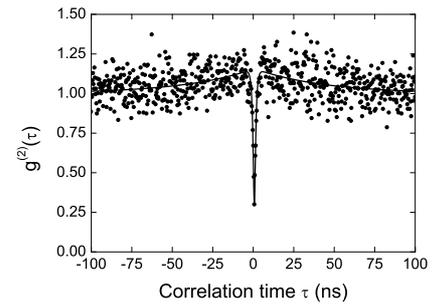}}
      \caption{Intensity correlation measurements of two color centers of Fig. \ref{spektren_vergleich} in the spectral
      window 765 -\SI{775}{nm}. The higher noise in the upper graph is due to the lower mean emission count rate of
      emitter (2) as compared to emitter (4).}
      \label{korrelationen}
\end{figure}

The sub-Poissonian character of the single Ni/Si centers' emission can be demonstrated by recording the second-order correlation function $g^{(2)}$. In Fig. \ref{korrelationen} the normalized intensity correlation measurements $g^{(2)}_{\text{meas}}(\tau)$ of color centers (2) and (4) of Fig. \ref{spektren_vergleich} are shown; the data have been corrected for background events corresponding to the signal-to-noise ratios of $\approx$\,6:1 for both emitters, a value which is obtained from the confocal scans. In both cases the value of $g^{(2)}_{\text{meas}}(0)$ is clearly smaller than 0.5, indicating the presence of single-photon emitters. For the derivation of the correct fit functions, we have to distinguish between the different emitters. As the correlation function of emitter (2) quickly reaches $g^{(2)}_{\text{meas}}(\tau)=1$ for $\tau>0$ and remains flat we assume a two level system -- in the equations below, the ground state is labeled as ``1'', the excited state as ``2''. Solving the rate equations for such a system yields
\begin{equation}\label{g2-funktion_2-level}
      g^{(2)}(\tau)=1-\text{e}^{-|\tau|/\tau_1}\,.
\end{equation}
The coefficient $\tau_1$ is defined as
\begin{equation}
      \tau_1=\frac{1}{r_{12}+r_{21}}\,,
\end{equation}
with  $r_{ij}$ representing the transition rate from state $i$ to state $j$. Because of the timing jitter of our HBT setup we have to convolute the function in Eqn. \ref{g2-funktion_2-level} with the device response function (DRF), measured to be a Gaussian peak with a $1/\sqrt{\text{e}}$ width of \SI{354}{ps}. Fitting the resulting function $g^{(2)}_{\text{fit}}$ to the data of emitter (2) yields a time constant of $\tau_1=\SI{1.45}{ns}$.

In case of center (4), $g^{(2)}_\text{meas}$ exceeds the value of one for time scales $\tau >\SI{5}{ns}$. This photon bunching is a sign for the presence of a longer lived shelving state (signed as ``3'') besides ground and excited state. According to Kitson et al. \cite{Kitson1998}, the $g^{(2)}$ function of such a three-level system is given by
\begin{equation}\label{g2-funktion_3-level}
      g^{(2)}(\tau)=1-(1+a)\,\text{e}^{-|\tau|/\tau_1}+a\,\text{e}^{-|\tau|/\tau_2}\,,
\end{equation}
with coefficients
\begin{equation}
      \begin{array}{c}
            \tau_{1,2}=2/(A\pm\sqrt{A^2-4B})\,,\\[0.5em]
            a=\frac{1-r_{31}\tau_2}{r_{31}(\tau_2-\tau_1)}\,,
      \end{array}
\end{equation}
and the abbreviations
\begin{equation}
      \begin{array}{c}
            A=r_{12}+r_{21}+r_{23}+r_{31}\,,\\[0.25em]
            B=r_{12}r_{23}+r_{12}r_{31}+r_{21}r_{31}+r_{23}r_{31}\,.
      \end{array}
\end{equation}
Eqn. \ref{g2-funktion_3-level} also has to be convoluted with the DRF to yield the correct fit function. The results for emitter (4) are $\tau_1=\SI{0.83}{}$ ns, $\tau_2=\SI{42.2}{ns}$, $a=0.16$.

The deviations of $g^{(2)}_{\text{fit}}(0)$ predicted by the fit functions and the measured values $g^{(2)}_{\text{meas}}(0)$ in both cases is smaller than 0.027. This small value proves that the difference of $g^{(2)}_{\text{meas}}(0)$ from zero is solely due to the limited timing resolution of our photon counting setup and the emitters show perfect anti-bunching.

In summary, the measured correlation functions indicate that emitter (2) behaves as a two-level system whereas emitter (4) has to be modeled as a three-level system. A similar distinction of emitters has recently been observed for Cr related color centers \cite{Aharonovich2009a}. We note here, that emitter (2) shows a spectrum that can be described as a single line, whereas the spectrum of emitter (4) is double-peaked with a separation of the two peaks of about \SI{3.6}{nm} (see Fig. \ref{spektren_vergleich}). The spectra of most investigated Si/Ni color centers fall in either one of these two categories.

\begin{figure}\centering
      \includegraphics[width=7.25cm]{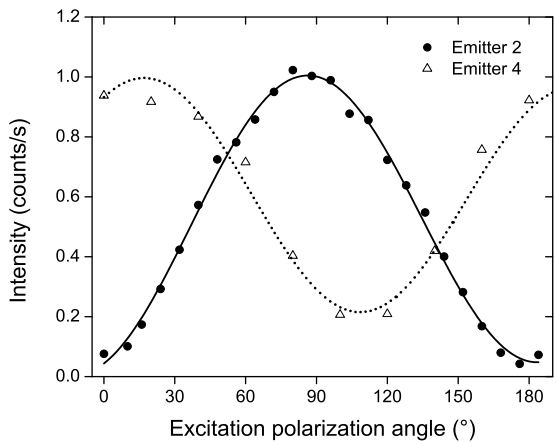}
      \caption{Count rates of emitters (2) and (4) in dependence on the excitation polarization. The fit functions are ordinary sines. The the curve of emitter (4) is shifted $\approx\SI{20}{\degree}$ due to a tilt of the sample in our setup.}
   \label{polarisation}
\end{figure}

To gain further insight into the Si/Ni center structure we investigated the polarization properties of the fluorescence. For emitter (2) we first measured the dependence of the count rate on the polarization angle of the excitation light (see Fig. \ref{polarisation}). It can be well described by a sinusoidal variation; the visibility
\begin{equation}
      V=\frac{I_{max}-I_{min}}{I_{max}+I_{min}}\,,
\end{equation}
amounts to \SI{92}{\%}, which means, that the excitation cross section is nearly a perfect dipole; the deviation from \SI{100}{\%} can be explained with depolarization introduced by the dichroic mirror in front of the objective and the objective itself. Fixing the excitation polarization to a maximum of the count rate and measuring the polarization degree of the light emitted from the color center yields a similar result: a periodicity of $\approx\SI{180}{\degree}$ with maxima at $\approx\SI{90}{\degree}$ and a polarization degree of \SI{98.5}{\%}. This observation is consistent with the existence of a single dipole. From the positions of the intensity maxima in both measurements and the known orientation of the fcc diamond lattice in our setup, one can deduce the alignment of the dipole or at least its projection along the $\langle 001\rangle$ direction.

On the other hand, emitter (4) shows a different behavior (see Fig. \ref{polarisation}): The periodicity of the count rate in dependence on the excitation polarization likewise is about $\SI{180}{\degree}$. The deviation of the angle of minimum count rate from $\SI{90}{\degree}$ can be explained with a certain tilt of the sample on the sample holder. However, the visibility amounts to \SI{62.1}{\%} only -- a fact that might be attributed to the presence of a second dipole at an orthogonal direction. A similar behavior is known from the NV center \cite{Davies1976} and for the \SI{1.4}{eV} Ni related center \cite{Collins1989}.

To summarize, one class of emitters (from now on labeled as ``2L'') behaves as two-level, single-dipole systems with a single emission line spectrum, whereas the other class (``3L'') exhibits a double-peaked spectrum, a three-level system and the presence of two orthogonal dipoles. These observations are consistent with a three-level model for the Si/Ni color center where ground state (``1'') and excited state (``2'') are connected via a fast radiative transition, and the metastable shelving state (``3'') lies close in energy to the excited state 2. For emitters of class 3L the splitting of levels 2 and 3 is pronounced, leading to the typical three-level behavior. On the other hand we might assume that for the emitters 2L local strain in the diamond crystal or a local charge distribution in the vicinity of the centers shift levels 2 and 3 into close degeneracy, such that populations can thermalize, resulting in an effective two-level system. The validity of our proposed model has to be verified by further experiments.
\begin{figure}\centering
      \includegraphics[width=8.5cm]{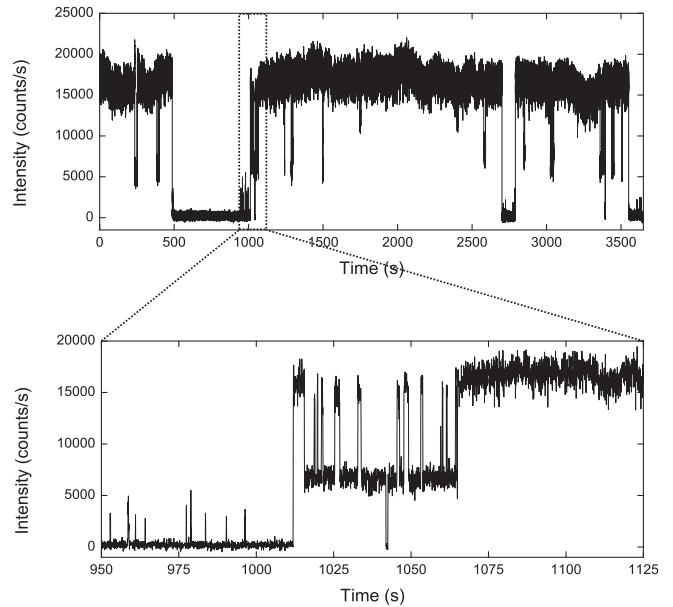}
      \caption{Mean count rate of emitter (2) in time windows of \SI{50}{ms} excited with an intensity of \SI{109}{kW/cm^2}
      recorded for about one hour. Additionally to the ``on'' and ``off'' level there is an intermediate level at $\approx\SI
      {6500}{counts/s}$.}
      \label{blinken}
\end{figure}

An important parameter with regard to a practical single photon source is the count rate of a single color center. The different emitters yielded different results: While the count rate of emitter (4) was absolutely stable, emitter (2) showed a fluctuating count rate. This blinking can clearly be seen in Fig. \ref{blinken} where the count rate is shown for a time of about one hour in time bins of \SI{50}{ms}. Besides periods of maximum count rate ($\approx\SI{17000}{counts/s}$) one can see ``off'' periods (\SI{0}{counts/s}) and periods, where the count rate reaches an intermediate value of $\approx\SI{6500}{counts/s}$. Such a kind of three-level blinking has been recently observed for luminescence from a semiconductor quantum dot \cite{Gomez2009}. In that investigation the existence of two emitting and one dark level has been explained by reversible electron trapping in surface states opening non-radiative decay channels. In case of the Ni/Si center an analogous process might be trapping or release of charges by impurity atoms or defects in the vicinity of the color centers. Here, the number of trapped charges or varying distances of trap sites to the color center would affect the radiative quantum yield differently. The blinking time distribution of the Ni/Si centers shows remarkably long dark intervals of up to \SI{500}{s}. Such long time scales are known from investigations on photochromism of NV color centers \cite{Iakoubovskii2000} and support the assumption of charge fluctuations. Anyway, the great length some dark periods renders a reliable usage of this color center as single photon source unfavorable.

\begin{figure}\centering
      \includegraphics[width=7.5cm]{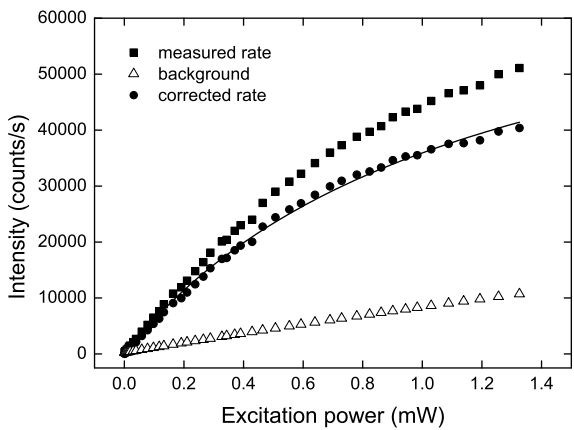}
      \caption{Saturation measurement of emitter (4). The background was recorded on the sample at a distance of \SI{10}{\micro m} from the color center. }\label{saettigung}
\end{figure}

On the other hand, for emitter (4) we did not observe blinking but stable and bright emission. A measurement of the saturation count rate is displayed in Fig. \ref{saettigung}. A fit for the count rate $R$ of the form
\begin{equation}
      R=R_\infty\frac{P}{P+P_\text{sat}}
\end{equation}
to the data yields a saturation power $P_\text{sat}$ of \SI{1.17}{mW} and a saturation count rate $R_{\infty}$ of \SI{77.8}{kcount/s}. This value is about \SI{50}{\%} lower than the \SI{200}{kcounts/s} reported for Ni/Si centers in nano-diamonds \cite{Aharonovich2009}, which can be explained with the better optical coupling efficiency in the case of nano-crystalline diamonds. Regarding the focus size of our setup we can estimate the saturation intensity as \SI{365}{kW/cm^2}.
For a rough estimate of the radiative excited state lifetime $\tau_f$ of the emitter we assume a two-level system. Following Trebbia et al. \cite{Trebbia2009} $\tau_f$ can then be calculated from $\tau_1$ by using the relation
\begin{equation}
      \tau_f=\tau_1\cdot\left(1+\frac{P}{P_\text{sat}}\right)=\SI{0.83}{ns}\cdot \SI{1.33}{} =\SI{1.11}{ns}\,,
\end{equation}
which takes into account the excitation power of the correlation measurement (\SI{390}{\micro W}) and the saturation power (\SI{1.17}{} mW).

The lifetime of Ni/Si related centers in diamond nano-crystals on a silica substrate was reported to amount to \SI{2}{ns} \cite{Aharonovich2009}. The deviation of our result to this value can be explained with the radiative lifetime of a single emitter placed in a dielectric material. This lifetime decreases linearly with growing refractive index $n$ \cite{Nienhuis1976}. In case of NV centers the radiative lifetime in bulk diamond ($n=\SI{2.4}{}$) was determined to be smaller by a factor of \SI{2.2}{} than in nano-crystals on silica \cite{Beveratos2001}. We here find a similar factor of \SI{1.8}{} for the ratio of Ni/Si color center lifetime in bulk diamond and nano-crystals. For the precise determination of the time constants excited state lifetime measurements (via pulsed laser excitation) or excitation power dependent $g^{(2)}$ measurements are required which is work in progress.

\subsection{Conclusion}

We report the observation of single Ni/Si related color centers in single crystalline diamond produced via ion implantation, marking an important step towards their deterministic production. With different dose combinations for both species we found maximal doses of \SI{E12}{cm^{-2}} for Ni and \SI{E10}{cm^{-2}} for Si to yield a color center density that is as high as possible and still keeps unwanted fluorescent background at a tolerable level. However, the creation efficiency is only about \SI{5E-8}{} in relation to the Ni dose and \SI{5e-6}{} in relation to the Si dose and our results indicate that it does not depend on the implantation energy. The Ni/Si center emission lines are centered within a spectral interval from 767 to \SI{775}{nm} with line widths between 1.4 and \SI{2.7}{nm} FWHM at room temperature. Single photon emission could be demonstrated with values for $g^{(2)}(0)$ that are clearly smal\-ler than 0.5. The offset of $g^{(2)}_{\text{meas}}(0)$ from zero can completely be explained with timing resolution of the HBT setup and the very short lifetimes of less than \SI{1.5}{ns} of the excited states of the emitters.

The investigated emitters can be divided into two groups: The emitters of the first group show a double peak structure in the emission spectrum, photon bunching in the correlation measurement and a polarization visibility that is clearly smaller than \SI{100}{\%} in excitation. The color centers of the second group exhibit only one peak in the spectrum, they do not show any bunching and they have a polarization contrast of $\approx\SI{98.5}{\%}$. For the explanation of this different behavior we propose a three level system with excited and shelving state lying close to each other in energy. Strain in the diamond lattice or local charge distributions can possibly lead to close degeneracy, making thermalization possible and introducing an effective two-level system.

For the Ni/Si centers we measured saturation count rates up to \SI{78}{kcounts/s}. Some of the centers showed strong three-level blinking which we tentatively explain by local charge fluctuations.

\section{Acknowledgements}

The ion implantations we were supported by Christian Hepp and Roland Albrecht (Universit\"at des Saarlandes, Saarbr\"u\-cken, Germany) and performed with Prof.\,Dr.\,Wolf\-gang Bol\-se (Universit\"at Stuttgart, Germany) at the Universit\"at G\"ottin\-gen, Germany and with PD\,Dr.\,Jan Meijer at the RUBION (Ruhr-Universit\"at Bochum, Germany). The project was financially supported by the Deutsche Forschungsgemeinschaft\linebreak and the Bundesministe\-rium f\"ur Bildung und Forschung (network EphQuaM, contract 01BL0903).

\end{document}